hep-lat/9509082   25 Sep 95

# Progress on lattice QCD algorithms

Philippe de Forcrand[a]

[a]Interdisciplinary Project Center for Supercomputing,
Swiss Federal Institute of Technology Zürich,
IPS, ETH-Zentrum, CH-8092 Zürich

I review recent progress on algorithms for calculating quark propagators and for simulating full QCD.

For the sake of brevity and consistency, I excluded from this review many interesting papers which discussed algorithms not directly relevant for QCD. I apologize to their authors, and refer the reader to their contributions in this volume.

The bottleneck of quenched lattice QCD is the calculation of quark propagators; for full QCD simulations, the Monte Carlo algorithm itself is the bottleneck. I discuss these two issues in Sections 1 and 2.

## 1. Quark propagator calculation

The task is to solve the linear system $A(U, m_q)x = b$, where $A(U, m_q)$ is the discretized Dirac operator. This task is to be repeated many times, over matrices $A$ constructed from different gauge fields $U$ with statistically similar properties, usually for a whole set of right-hand sides $b$, and often for a whole range of quark masses $m_q$. Let us first consider the solution of a single system.

The matrix $A$ is not hermitian, so that the standard Conjugate Gradient (CG) algorithm can only be applied to the system $A^\dagger Ax = A^\dagger b$ (it is then abbreviated CGNE: CG on the Normal Equations). CGNE is expected to have mediocre convergence properties, because the condition number of $A^\dagger A$ is the square of that of $A$. Therefore, until recently, the preferred method was nearly the simplest: do a line minimization of $\| Ax - b \|$ at each iteration, in the search direction defined by $A$ times the previous residual $r \equiv Ax - b$, but orthogonal to the previous search direction. This algorithm, called Conjugate Residual (CR) [also called CR(1) or GM-RES(1)] offers no minimum rate of convergence, and is therefore supplemented by CGNE: one switches to CGNE when convergence of CR becomes unsatisfactory.

The situation has changed with the recognition that there exist other iterative solvers, designed for non-hermitian matrices, which converge faster than CR. Early work on this subject [1,3] stressed the advantages of BiCG-type methods, and quantified the gain over CR and CG: the number of matrix-vector products is reduced by a factor $\sim 2$. The question then arises: can one do better ?

It is possible to answer this question for all current algorithms because they are all Krylov methods. By that one means that at each iteration $k$, an approximate solution $x_k$ is found in the vector space (Krylov space)

$$\mathcal{E}_k \equiv span\{r_0, Ar_0, A^2r_0, ..., A^{k-1}r_0\} \qquad (1)$$

Different algorithms come more or less close to finding in $\mathcal{E}_k$ the vector $\tilde{x}_k$ which minimizes the norm of the residual $\| Ax - b \|$. An exact determination of $\tilde{x}_k$ requires the construction of an orthonormal basis of $\mathcal{E}_k$. When $A$ is hermitian positive definite, CG has the remarkable property of building this basis recursively, with only 2 inner products per iteration, and finding $\tilde{x}_k$ exactly: it can be called an "optimal" algorithm, in the sense that the number $k$ of matrix-vector products required to reduce the norm of the residual to a given tolerance is the minimum possible. When $A$ is non-hermitian, the progressive construction of an orthonormal basis of $\mathcal{E}_k$ for finding $\tilde{x}_k$ necessitates the evaluation of $k$ inner products $< \vec{e}_i | \vec{e}_j >$ at each iteration $k$, and the storage of $k$ vectors: the work grows like $k^2$ and the memory requirement like $k$. These two demands render this algorithm, called (full) GMRES (Generalized Minimum RESidual) impractical except for tests.

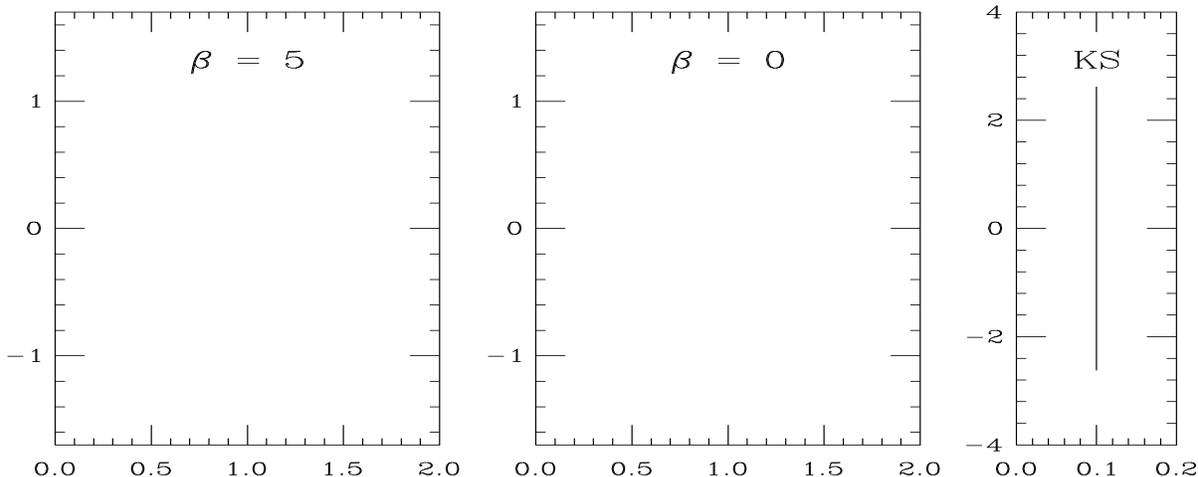

Figure 1: Typical eigenvalue spectra for the Dirac matrix $A/2$, for Wilson fermions at $\beta = 5$ and $\beta = 0$, and for staggered (KS) fermions

But GMRES will achieve for non-hermitian matrices $A$ what CG did in the hermitian case: it will provide a lower bound on the number of matrix-vector products necessary to reach the stopping criterion. We can then assess how far from optimal other algorithms are, by comparison with this lower bound. We do this in turn for Wilson and Staggered fermions, then review directions for further progress.

### 1.1. Wilson fermions

The matrix is $A \equiv \mathbf{1} - \kappa M$, with symmetries:
• $\Sigma M \Sigma = -M$, where $\Sigma = \pm 1$ on even/odd lattice sites. This is because $M$ is a hopping matrix, connecting nearest neighbours on the lattice ($M$ has a block-block-block-tridiagonal structure).
• $\gamma_5 M \gamma_5 = M^\dagger$. This is inherited from the continuum operator.
These symmetries respectively imply that eigenvalues of $M$ come in opposite pairs, and are real or come in complex conjugate pairs (real eigenvalues only appear in the confined phase). A typical spectrum of $A/2$ is shown in Fig.1a (confined phase); as $\beta$ decreases to 0, fluctuations in $A$ increase and the spectral density of $M$ becomes uniform in the disk of radius 4 centered at the origin (see Fig.1b, where the spectrum of $A/2$ is shown).

A comparison of algorithms at $\beta = 0$ (Fig.2) shows the advantages of non-hermitian solvers, in particular those based on the Bi-Conjugate Gradient (BiCG, [6]). BiCG constructs 2 Krylov spaces, $\mathcal{E}_k$ (1) and

$$\tilde{\mathcal{E}}_k \equiv span\{s_0, A^\dagger s_0, A^{\dagger 2} s_0, ..., A^{\dagger k-1} s_0\} \quad (2)$$

In these 2 spaces, sequences $\{u_i\}$ and $\{v_j\}$ which are mutually orthogonal are built: $< u_k | v_l > = 0 \ \forall \ l \leq k$. This orthogonality property ensures convergence after $n$ steps, $n$ being the rank of the matrix, in exact arithmetic, although convergence need not be monotonic. But it also contains the germ of numerical instabil-

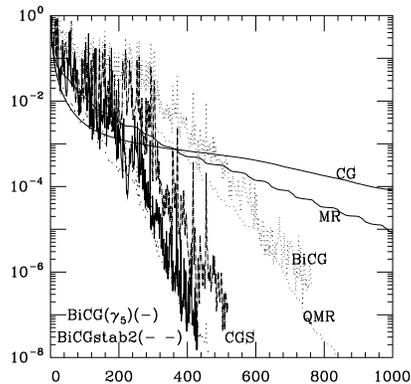

Figure 2. Norm of the residual versus number of matrix-vector products, for various solvers at $\beta = 0, \kappa = .2475$ ($4^4$ lattice).

ity: the vanishing of $< u_k | v_l >$ is the result of many cancellations; and $x_k$, which is built from a linear combination of $Au_k, u_k$, and $u_{k-1}$, can have large numerical round-off errors. The situations of breakdown (division by zero - see below) or near-breakdown may occur, and "look-ahead" or "stabilized" BiCG algorithms are designed to handle these difficulties gracefully.

However, one can try the simplest algorithm, which uses the $\gamma_5$-symmetry of $M$ to obtain the sequence $\{v_j\}$ for free, by just taking $s_0 = \gamma_5 r_0$. In pseudo-code, this algorithm, which we call BiCG$\gamma_5$, reads

*Choose starting solution $x_0$.*
$r_0 = b - A * x_0$
$p_0 = r_0$
$\delta_0 = < r_0 | \gamma_5 | r_0 >$
*for $n = 0, 1, \ldots$ until convergence*
$\omega_n = \delta_n / < p_n | \gamma_5 A | p_n >$
$x_{n+1} = x_n + \omega_n * p_n$
$r_{n+1} = r_n - \omega_n * A * p_n$
$\delta_{n+1} = < r_{n+1} | \gamma_5 | r_{n+1} >$
$p_{n+1} = r_{n+1} + \delta_{n+1}/\delta_n * p_n$
*end for*

It is identical to CG, except that a $\gamma_5$ has been inserted in the inner products. The possibility of breakdown comes in the division by $< p_n | \gamma_5 A | p_n >$ and by $\delta_n$. We have experienced no such breakdown in solving several $10^4$ systems, on a 64-bit Cray computer, with a maximum lattice size $8^3 \times 16$; but the reader who uses a 32-bit machine would be well-advised to use a stabilized variant [4]. As Fig.2 shows, BiCG$\gamma_5$ compares equally with BiCGStab2, our best stabilized BiCG algorithm, but requires fewer storage vectors and fewer inner products. We like it mostly for its simplicity.

Comparing BiCG$\gamma_5$ to GMRES as a function of $\kappa$ in Fig.3 reveals that the number of matrix-vector products required by BiCG$\gamma_5$ is within $\sim 10\%$ of the absolute minimum, over a comprehensive range of quark masses.[1]

---

[1] Admittedly, the comparison should be repeated on an ensemble of gauge configurations larger than $8^3 \times 16$. The cost of GMRES has so far prevented us from doing so.

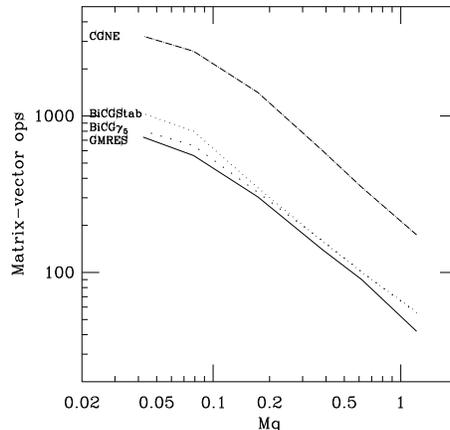

Figure 3. Number of matrix-vector products needed to reduce the norm of the residual by $10^{10}$, for various solvers, as $\kappa$ is varied. The quark mass $m_q$ is estimated as $(\kappa^{-1} - \kappa_c^{-1})/2$, with $\kappa_c = .1694$. The lattice size is $8^3 \times 16$; $\beta = 5.7$.

In my opinion, it is not worth looking for a better algorithm: BiCG, in its stabilized or in its $\gamma_5$ version, is quasi-optimal.

## 1.2. Staggered fermions

In this case, the matrix is $A \equiv m\mathbf{1} + iB$, where
- $\Sigma B \Sigma = -B$ as for Wilson fermions;
- $B^\dagger = B$.

Thus the eigenvalues of $B$ are real and come in opposite pairs. The spectrum of $A$ lies on a vertical line segment (Fig.1c).

We compared GMRES with CGNE. They require the same number of matrix-vector products, exactly proportional to $1/m$. Thus CGNE is optimal, confirming the common wisdom that nothing can be done to improve convergence over CGNE. The question is why.

A theorem by Voevodin [5] states that the orthogonalization of the basis in GMRES, which normally requires $\mathcal{O}(k)$ inner products at the $k^{th}$ iteration, can be reduced to a short recurrence and $\mathcal{O}(1)$ inner products whenever matrix $A$ is of the form

$$A = e^{i\theta}(B + \sigma\mathbf{1}), \quad \theta \in \mathcal{R}, \quad \sigma \in \mathcal{C} \quad (3)$$

Therefore a CG-like algorithm must exist for staggered fermions. Why is it exactly CGNE ?

All Krylov methods effectively build at itera-tion $k$ a polynomial $P_k(A)$ in the matrix $A$, such that $P_k(A) \approx A^{-1}$, and apply it to the right-hand side $b$. The polynomial built by GMRES, $P_k(A)$ or equivalently $Q_k(B)$, is optimal; and because the spectrum of $B$ is even, the polynomial $Q_k(B)$ will also be even. CGNE builds a polynomial in $A^\dagger A \equiv m^2 \mathbf{1} + B^2$, i.e. it is an even polynomial in $B$ by construction; and it is the optimal polynomial in $B^2$ because CG is optimal for a hermitian positive system. Therefore the 2 polynomials built by GMRES and CGNE are both even in $B$ and optimal: they are identical.

Thus optimal and near-optimal solvers are available for staggered and Wilson fermions respectively. The next improvement should come from preconditioners: but it should be clear that polynomial-type preconditioners will not reduce the amount of work needed, since they do not extend the search for a solution outside the original Krylov space, which is already explored almost perfectly.

## 1.3. Further progress

Further gains can be achieved by avoiding redundant work during the repetitive solution of related systems. Repetition over masses, right-hand sides, and gauge configurations are discussed in this order.

### 1.3.1. Multiple masses

The structure of matrix $A$ lends itself to an economical calculation of several systems

$$(m_i \mathbf{1} + iB)x_i = b, \quad i = 1, ..., m$$

(using staggered fermions to be specific), because the Krylov space $\mathcal{E}_k$ is invariant under a change of the quark mass $m_i$. It is sufficient to apply the Lanczos process to matrix $B$ once, and then to recombine the Lanczos vectors for any set of masses. Only one matrix-vector product per iteration is necessary. The savings are machine-dependent, but can be considerable, well beyond those obtained by the usual procedure of taking the last solution as an initial guess for the next lighter quark mass. The price to pay is the storage of all Lanczos vectors of B on disk (for an *a posteriori* reconstruction of other propagators for any mass), or of an extra pair of Lanczos vectors per mass value in memory (for on-the-fly re-construction). Even-odd preconditioning can be preserved at an overall cost of 2 in the above.

Details have been well explained in [2]. This simple idea should greatly improve the status of quark mass interpolations in quenched QCD.

### 1.3.2. Multiple right-hand sides

Hadron propagators are built from quark propagators with sources (right-hand sides) spanning the color and Dirac spaces. Sources having different $x$-space profiles are often desired for variational purposes. Thus one is lead to solving

$Ax_i = b_i, \quad i = 1, ..., m$

with $m \sim \mathcal{O}(12)$ or more.

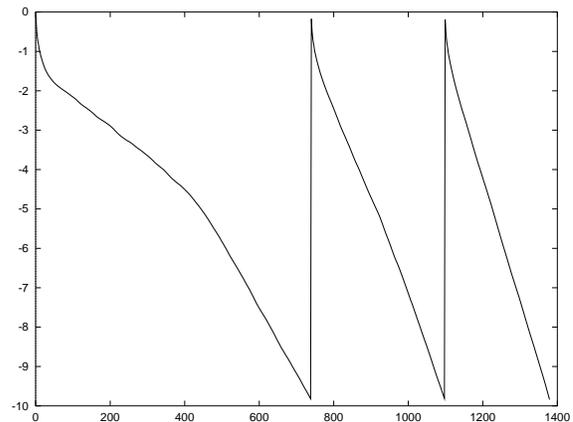

Figure 4. $\text{Log}_{10}$ of the norm of the residual, as a function of the number of matrix-vector products, for multiple right-hand sides, showing how the system "learns" to converge faster. The gauge configuration is the same as in Fig.3; $\kappa = .167$.

This is the subject of intense work in the numerical analysis community [7]. Let me report some preliminary results [8]. The idea is to keep in memory the most significant search directions as a rectangular matrix $Q$, and then to "deflate" the system by solving $(\mathbf{1} - QQ^\dagger)Ax = (\mathbf{1} - QQ^\dagger)b$. A nested process allows the system to progressively refine its set of vectors $Q$, and to "learn" as it iterates. Only one right-hand side is considered at a time. Fig.4 shows the history of the norm of the residual as a function of the number $n$ of matrix-vector products, for 3 sources on different lattice sites. On the second r.h.s., the system has learned rather well the significant directions which slow down convergence, and progresses a little more on the third system. Nonetheless, the reduction in $n$ is only $\mathcal{O}(2)$. It will increase as the right-hand sides become more linearly dependent.

### 1.3.3. Multiple configurations

The eigenvalue spectrum of matrix $A$, overall, varies little from one configuration $\{U\}$ to another at the same $\beta$. In fact, it varies little also as one scales the size of the lattice: the average density of eigenvalues in some region of the complex plane essentially scales like the volume of the lattice. Thus valuable information on the eigenvalue spectrum, and possibly the structure of the eigenvectors after gauge-fixing, could be obtained from a small lattice study, e.g. for the purpose of building a preconditioner.

## 2. Full QCD Monte Carlo

### 2.1. Extrapolation methods

Rather than facing the cost of a full QCD simulation, one may consider cheaper approximations which extrapolate to the full QCD result. The simplest extrapolation is in $n_f$, the number of flavors. This approach, considered a long time ago [9], has been revived under the name of "bermions" [10]. Consider adding to the gauge action a term $\phi^\dagger(\mathbf{1} - \kappa M)^\dagger(\mathbf{1} - \kappa M)\phi$, ie. a bosonic term with fermion-like interaction (hence the name). Integration over $\phi$ gives $det^{-2}(\mathbf{1} - \kappa M)$, as would result from -2 quark flavors. Additional pairs of bermions can, if desired, mimic $n_f = -4, -6, \ldots$. Together with the quenched result $n_f = 0$, one can extrapolate to $n_f > 0$. A linear extrapolation appears surprisingly good for heavy quarks, when compared with full $n_f = 2$ results. Bermions are attractive because they are cheap to simulate; the systematic error in the $n_f$ extrapolation is unclear.

Another strategy proposed in [12] consists in applying to quenched results an estimate of the correction due to unquenching. The leading effect of dynamical fermions is a renormalization of the gauge coupling, which can be incorporated in the quenched Monte Carlo at no cost. The next-order correction is then computed for each observable.

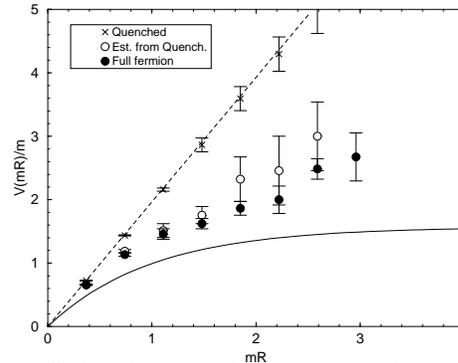

Figure 5. Static potential in the Schwinger model. The white circles are obtained from quenched results corrected for unquenching. From [11].

The results are remarkable: the remaining difference with full dynamical quarks is buried in the statistical noise. This is illustrated in Fig.5, taken from [11], for the static potential in the Schwinger model. This is yet another confirmation that dynamical fermions generate an essentially local effective action, unless they are very light. It makes one wonder anew about possibilities of guiding a full QCD simulation with a short-range effective action, and enforcing correct sampling with a Metropolis test: the acceptance will fall exponentially with the volume but, on the other hand, the autocorrelation time will be reduced by a large factor, and probably also a power of the volume, over Hybrid Monte Carlo. For moderate volumes and quark masses, this simple strategy may win.

### 2.2. Hybrid Monte Carlo

A good deal of progress in integrating the equations of motion along an HMC trajectory should allow, when combined with the non-hermitian solvers of section 1.1, a gain $\mathcal{O}(10)$ in efficiency. Here are the main ingredients:

1) Introduce different "time steps" when computing the force coming from the gauge fields and that coming from the external boson fields [13]. The former can be evaluated cheaply at frequent intervals. This will reduce the overall discretization error in the molecular dynamics integration. Ref. [14] claims savings by a factor $\sim 2$.

2) Use previous solutions of $Ax = b$ at earlier time-steps on the trajectory. The simplest, widely used extrapolation, consists in choosing as a starting guess at step $i$: $x_{0i} = 2x_{i-1} - x_{i-2}$. A more efficient approach [15] minimizes $\| Ax - b \|$ over the span of the earlier solutions, achieving another savings $\sim 2$. My understanding is that the system $Ax = b$ could be, in addition, "deflated" from the directions already explored, leading to faster convergence still.

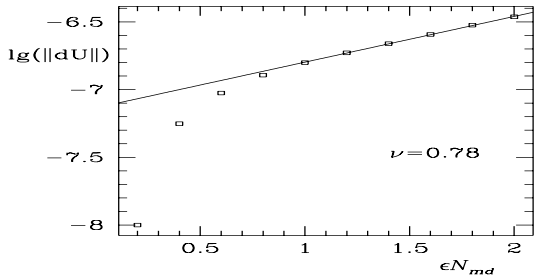

Figure 6. Logarithm of the change in the gauge field caused by a small noise at time 0, versus time. From [14].

3) Use an adaptive time step. This is a very promising development, since HMC trajectories at low quark mass tend to bounce off energy barriers caused by a small determinant separating different topological sectors. It would make sense to reduce the step size during those "sharp curves", and accelerate afterwards. Paradoxically, it is possible to implement a variable step-size while preserving reversibility of the molecular dynamics evolution [16]. Consider an elementary evolution operator $E_\tau \equiv e^{A(q)\tau/2} e^{B(p)\tau} e^{A(q)\tau/2}$, where $A$ and $B$ act on conjugate variables $p$ and $q$, and $\tau$ is the step size. $\tau$ can be varied adaptively as a function of some error $e(\tau, \{p,q\}(t))$ provided
• the error function is symmetrized, ie.
$e_S(\tau, \{p, q\}(t)) =$
$e(\tau, \{p, q\}(t)) + e(-\tau, E_\tau(\{p, q\}(t)));$
• at each step the bound on the error must be saturated. Thus $\tau$ is the solution of a nonlinear equation $e_S(\tau, \{p, q\}(t)) = tolerance$.
These 2 conditions guarantee that the trajectory can be retraced step by step, by reversing the momenta. This algorithm, tested on the Kepler problem with spectacular results, awaits implementation for QCD.

In spite of these promising developments, caution is required before jumping to larger lattices and smaller quark masses. Roundoff errors become sizeable on global sums and inner products, especially on 32-bit machines. The molecular dynamics integration amplifies these errors exponentially, as best shown in [14] who measured the Lyapunov exponent characteristic of this chaotic behavior (see Fig.6). HMC relies on long trajectories, $\sim 1/m$, where $m$ is the smallest mass in the system, to maintain a dynamical exponent $z = 1$, and on reversibility to converge to the correct distribution. Roundoff errors introduce dissipation and spoil reversibility. For current trajectories of $\sim 100$ steps or more, at the smallest quark masses and on the largest lattices, lack of reversibility becomes a serious problem [17].

## 2.3. Kramer's algorithm

This algorithm, proposed by Horowitz [18] and also known as L2MC [19], is a variant of HMC based on the second-order Langevin equation:
$\dot{q} = \partial H/\partial p$
$\dot{p} = -\partial H/\partial q - \gamma p + \sqrt{2\gamma}\eta$
where $\gamma$ is the viscosity coefficient and $\eta$ a Gaussian noise. The guiding trajectory of HMC now includes some dissipation. The change to the usual leapfrog integration scheme is trivial: one intercalates an irreversible step $p \leftarrow e^{-\gamma\tau}p + \sqrt{1 - e^{-2\gamma\tau}}\eta$. In addition, all momenta must be reversed after rejection by the Metropolis test.

The original motivation was that the extra tunable parameter $\gamma$ could be adjusted to accelerate the dynamics. The optimal value of $\gamma$ has been studied for free field in [19]: not surprisingly, $\gamma_{opt} \sim 1/m$, so that momenta are effectively refreshed after a time $\sim 1/m$ as in HMC. Neither algorithm appears clearly more efficient [14]. Kramer's algorithm, however, maintains a dynamical exponent $z = 1$ for arbitrary short trajectories.

I see two advantages in favor of Kramer's algorithm:
• Since the dynamics is intrinsically dissipative, the additional dissipation due to roundoff errors is negligible. Problems of irreversibility unearthed with HMC can safely be ignored.
• By making trajectories shorter, more configurations will be generated for the same computer cost. They will not be completely independent, but for some observables with short integrated autocorrelation time (like glueballs), they effectively will be.

## 2.4. Lüscher's method

Lüscher's original proposal [20] has now been studied thoroughly; it has been modified and improved a great deal. The two ingredients are:

• approximate $det(A)$ by $det^{-1}(P(A))$, where $z\ P(z) \approx 1$ over the whole spectrum of $A$ (see Fig.1, with $\sigma_{min}(A) \approx m_q$)

• factorize $P(A) \propto \prod_k (A - z_k)$, group the factors into positive pairs, and express each as a bosonic Gaussian integral. Thus

$$det(A) \approx \int \prod_k \mathcal{D}\phi_k^\dagger \mathcal{D}\phi_k e^{- \sum_k \phi_k^\dagger (A - z_k)^\dagger (A - z_k)\phi_k} \quad (4)$$

The approximation is controlled by the degree $n$ of $P$ (ie. the number of bosonic species), and converges *exponentially*:

$|\lambda\ P(\lambda) - 1| \le c_1\ e^{-c_2\ n\ /\ m_q}$,

with $c_1, c_2 \sim \mathcal{O}(1)$, for all eigenvalues $\lambda$ of $A$. Therefore it is sufficient to increase $n$ as $m_q^{-1} Log V$ with the quark mass $m_q$ and the volume $V$ of the lattice to keep the quality of the approximation constant.

Early results showed disappointingly slow Monte Carlo dynamics [21]. In fact one observes that, if the bosonic fields $\phi_k$ are frozen, the gauge fields decorrelate very quickly. Thus the slow dynamics are governed by the correlation length of the bosonic terms. Its maximum is $\mathcal{O}(m_q^{-1})$. Calling $z$ the dynamical exponent of the bosonic Monte Carlo, one gets for the complexity of the algorithm:

Nb. of bosonic fields $\propto m_q^{-1}$

Autocorrelation time $\propto m_q^{-z}$

Work $\propto m_q^{-1-z}$

Although over-relaxation has been used, all results are consistent with $z = 2$. Since the bosonic action is a simple Gaussian, there is some hope that cluster or multigrid Monte Carlo will provide a reduction in $z$. Even with $z = 2$, Lüscher's algorithm behaves better asymptotically than Hybrid Monte Carlo, where the work grows at least as $V^{5/4} m_q^{-13/4}$ [22].

A large reduction in the work and in the number of bosonic fields comes by implementing for $P(A)$ the same ideas which have proven successful for the quark propagator calculation: even-odd preconditioning [23], non-hermitian variant [24] (which also allows the simulation of an odd

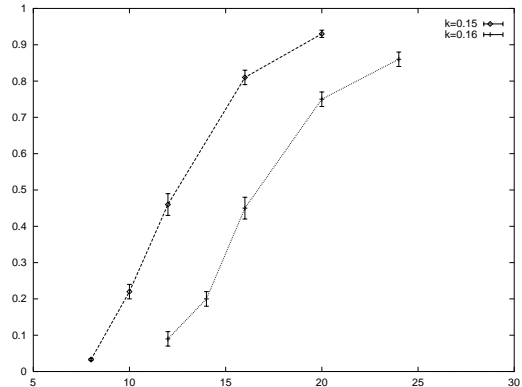

Figure 7. Metropolis acceptance versus number of bosonic fields in the exact, non-hermitian Lüscher algorithm; the lattice is $4^4$; $\beta = 5$. From [29].

number of flavors).

Most significant, however, has been the suggestion to make the algorithm exact with the adjunction of a Metropolis test [25]. The acceptance for a transition $U \to U'$ should then be

$$min(1, \frac{det^{-2}A(U')P(A(U'))}{det^{-2}A(U)P(A(U))})$$

Early efforts aimed at evaluating the determinant ratio exactly with a Lanczos process [25,27]. This approach was plagued by roundoff errors and by a cost scaling like $V^2$. Instead, it is sufficient to estimate this ratio stochastically [24,26] with one Gaussian vector $\eta$ as $e^{-\eta^\dagger (W^\dagger W - 1)\eta}$ with $W \equiv [A(U')P(A(U'))]^{-1} A(U)P(A(U))$. A proof of detailed balance is given in [28]. Each Metropolis test then entails the solution of a linear system; one can show that the associated cost, measured in update sweeps, remains constant as $V$ and $m_q^{-1}$ increase.

As an example of the results currently obtained with this algorithm [29], Fig.7 shows the Metropolis acceptance as the number of fields $n$ is varied, for 2 different quark masses.

A comparison of efficiency with HMC must be done carefully. Fig.8 shows some very preliminary data, presented on a scale which attempts to remove volume effects. Work is measured in matrix-vector products per independent configuration; for lack of better data, configurations for which the plaquette is decorrelated are considered independent. The impression I want to convey from this figure is that the exact, non-hermitian

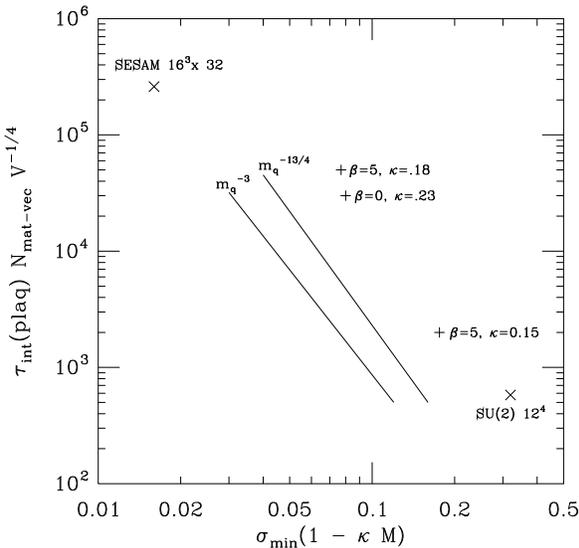

Figure 8. The number of matrix-vector products necessary to decorrelate the plaquette, divided by $V^{1/4}$, is shown as a function of the lowest singular value of the Wilson matrix, for HMC (X) [17,14] and Lüscher (+) [29] simulations.

variant of the Lüscher algorithm is a reasonable alternative to HMC.

## 3. Acknowledgements

I thank A. Boriçi, A. Borrelli, R. Brower, R. Edwards, R. Freund, A. Frommer, A. Galli, U. Heller, A. Kennedy, T. Lippert, P. Millard, M. Peardon, K. Ressel, E. de Stürler and D. Weingarten for help and comments.